\documentstyle[12pt,psfig]{article}

\begin{document}

\begin{titlepage}
\pagestyle{empty}
\rightline{LBL-37234}
\rightline{hep-th/yymmddd}
\rightline{May 1995}
\vskip .2in
\baselineskip=18pt
\begin{center}
{\large{\bf Studying Non-calculable Models of\\ Dynamical Supersymmetry
Breaking}
}\footnote{This work was supported by the
Director, Office of Energy Research, Office of High Energy and Nuclear
Physics, Division of High Energy Physics of the U.S. Department of
Energy
under Contract DE-AC03-76SF00098.}
\end{center}

\vskip .1in
\begin{center}

Hitoshi Murayama\footnote{On leave of absence from {\it Department of
Physics, Tohoku University, Sendai, 980 Japan}}

{\it Theoretical Physics Group, Lawrence Berkeley Laboratory}

{\it University of California, Berkeley, CA 94720, USA}

\vskip .1in

\end{center}
\vskip .2in
\centerline{ {\bf Abstract} }
\baselineskip=18pt
There are supersymmetric gauge theories which do not possess any
parameters nor flat directions, and hence cannot be studied anywhere in
the field space using holomorphy (``non-calculable'').  Some of them are
believed to break supersymmetry dynamically.  We propose a simple
technique to analyze these models.  Introducing a vector-like field into
the model, one finds flat directions where one can study the dynamics.
We unambiguously show that the supersymmetry is broken when the mass of
the vector-like field is small but finite, and hence Witten index
vanishes.  If we increase the mass of the vector-like field, it
eventually decouples from the dynamics and the models reduce to the
original non-calculable models.  Assuming the continuity of the Witten
index in the parameter space, one can establish the dynamical
supersymmetry breaking in the non-calculable models.

\noindent
\end{titlepage}
\renewcommand{\thepage}{\roman{page}}
\setcounter{page}{2}
\mbox{ }

\vskip 1in

\begin{center}
{\bf Disclaimer}
\end{center}

\vskip .2in

\begin{scriptsize}
\begin{quotation}
This document was prepared as an account of work sponsored by the United
States Government. While this document is believed to contain correct
information, neither the United States Government nor any agency
thereof, nor The Regents of the University of California, nor any of their
employees, makes any warranty, express or implied, or assumes any legal
liability or responsibility for the accuracy, completeness, or usefulness
of any information, apparatus, product, or process disclosed, or represents
that its use would not infringe privately owned rights.  Reference herein
to any specific commercial products process, or service by its trade name,
trademark, manufacturer, or otherwise, does not necessarily constitute or
imply its endorsement, recommendation, or favoring by the United States
Government or any agency thereof, or The Regents of the University of
California.  The views and opinions of authors expressed herein do not
necessarily state or reflect those of the United States Government or any
agency thereof, or The Regents of the University of California.
\end{quotation}
\end{scriptsize}

\vskip 2in

\begin{center}
\begin{small}
{\it Lawrence Berkeley Laboratory is an equal opportunity employer.}
\end{small}
\end{center}

\newpage
\renewcommand{\thepage}{\arabic{page}}
\setcounter{page}{1}

Supersymmetry is an attractive possibility to stabilize
the hierarchy between the weak- and unification- or Planck-scales.
Especially interesting is the case where the
electroweak symmetry cannot be broken in the supersymmetric limit, such
as in the minimal supersymmetric standard model, since
one can understand the smallness of the weak scale in terms of the
smallness of the supersymmetry breaking effects.  However, the origin of
the hierarchy itself remains unexplained in lack of
understanding why supersymmetry is (weakly) broken.
Dynamical supersymmetry breaking is a natural idea to explain the
smallness of the supersymmetry breaking scale \cite{Witten}.  Indeed,
a class of chiral gauge theories were shown to break supersymmetry
dynamically \cite{ADS}, and can be used to construct realistic models
\cite{Nelson}.  Since then, there was a substantial progress in
the technique to analyze dynamics of supersymmetric gauge theories based
on holomorphy \cite{Seiberg,NS,exact,more}.  The technique was also applied
to build new models which break supersymmetry dynamically
\cite{recentDSB}.

The earliest models of the dynamical supersymmetry breaking
\cite{SU5,SO10}, however, cannot be analyzed using the holomorphy.
The known examples are SU(5) theory with {\bf 5$^*$} and {\bf 10}, and
SO(10) with a single {\bf 16}.\footnote{There are also models with
non-abelian flavor symmetries \cite{ADS} even though we do not discuss
them in this letter.  The framework in \cite{LMM} should be useful to
analyze such models.} These models do not have any adjustable parameters
nor any flat directions, and hence ``non-calculable''.  They were argued
to break supersymmetry dynamically because of the following reason.  These
theories possess an U(1)$_R$ symmetry.  If the low energy theory
preserves U(1)$_R$, the low energy particle content should saturate the
anomalies of the fundamental theories.  There are possible candidates of
such low energy particle contents.  But it was argued such particle
contents are ``implausible'' because of the complicated charge
assignments.  Then it is more ``plausible'' to have U(1)$_R$ symmetry
spontaneously broken, and one needs its non-linear realization.
However, it tends to require a flat direction in the low-energy theory.
This is also argued to be ``implausible'' since the fundamental theory
did not possess any flat directions.  Even though a
strong case was made, it is still desired to have a method to analyze
these models where one can explicitly see the breakdown of
supersymmetry.

The purpose of this letter is to point out there is a simple method to
study the ``non-calculable'' models by introducing additional
vector-like field (field which transforms under a real representation
of the gauge group) to the models.  The original models are understood as
the limit where the vector-like fields decouple.  When the vector-like
field is massless, there are flat directions and the models can be
analyzed using by-now well-known technique of holomorphy.  Once we turn
on the mass of the vector-like field, the models break supersymmetry
spontaneously, and hence Witten index vanishes.  Assuming the continuity
of the phase as we increase the mass of the vector-like field, Witten
index vanishes in the limit where the vector-like fields decouple.
There is no sign of supersymmetry restoration when one gradually raises
the mass of the vector-like field.  Then one can conclude that the
original models break supersymmetry dynamically.

\begin{table}
\centerline{
\begin{tabular}{c|cc|cc|cc}
& $H$ & $\psi$ & $\psi\psi H$ & $H^2$ & $\lambda$ & $M$ \\ \hline
U(1)$_R$ & 1 & $-$3 & $-$5 & 2 & 7 & 0\\
U(1)$_M$ & 2 & $-$1 & 0 & 4 & 0 & $-$4
\end{tabular}
}
\caption{Charges of the fields and parameters under non-anomalous global
symmetries in the SO(10) model with $\psi$({\bf 16}) and $H$({\bf 10}).}
\end{table}

We discuss an SO(10) model with a single $\psi$({\bf 16}).  It was shown
that this model does not have any flat directions \cite{SO10}.  We
introduce a vector-like field $H$ which transforms as a {\bf 10}.  The
non-anomalous global symmetries of the model are listed in Table~1.
There are two non-anomalous symmetries in this model, an $R$-symmetry
U(1)$_R$ and a non-$R$ symmetry U(1)$_M$.  In the absence of the
superpotential, the most general flat direction of this model (up to
gauge transformations) is parametrized by three complex scalar fields
$H^\pm$ and $\chi$,
\begin{eqnarray}
&&H^1 = \cdots = H^8 = 0, \\
&&H^9 = \frac{i}{\sqrt{2}} (H^+ - H^-), \\
&&H^{10} = \frac{1}{\sqrt{2}} (H^+ + H^-), \\
&&\psi = (\uparrow \otimes \uparrow \otimes \uparrow \otimes \uparrow) \chi,
\end{eqnarray}
with the $D$-flatness condition
\begin{equation}
|H^+|^2 - |H^-|^2 - \frac{1}{2} |\chi|^2 = 0. \label{D-flat}
\end{equation}
See appendix for notation.  The low-energy theory is a pure
SO(7) supersymmetric Yang--Mills theory with two singlet chiral
superfields,\footnote{When SO(10) breaks down to SO(7), $45 - 21 = 24$ chiral
superfields are eaten.  It leaves $16 + 10 - 24 = 2$ chiral superfields
massless.} which can be identified with gauge-invariant
composite fields $\psi \psi H$ and $H^2$ (the gauge indices are contracted in
an obvious manner).  There is a unique possible superpotential generated
non-pertubatively by the condensate of SO(7) gauginos \cite{ADS2},
\begin{equation}
W_{\rm n.p.} = c \frac{\Lambda^{21/5}}{(\psi \psi H)^{2/5}} \ .
\end{equation}
The coefficient $c$ is a constant of order unity.  There is no ground
state in the absence of a tree-level potential.

There are two possible terms in the superpotential at the tree-level,
\begin{equation}
W_{\rm tree} = \lambda \psi \psi H + \frac{1}{2} M H^2 .
\end{equation}
Charges of the parameters $\lambda$ and $M$ under the global symmetries
are also listed in Table~1.  The total superpotential is the sum of
$W_{\rm tree}$ and $W_{\rm n.p.}$ and is exact
in the sense of \cite{exact} as shown below.  Because of the U(1)$_R$
and U(1)$_M$ symmetries, the superpotential has to take the form
\begin{equation}
W_{\rm total} = c \frac{\Lambda^{21/5}}{(\psi \psi H)^{2/5}} \
	F \left( \frac{\lambda \psi \psi H}
		{c\Lambda^{21/5}/(\psi \psi H)^{2/5}} ,
		\frac{M H^2/2}{c\Lambda^{21/5}/(\psi \psi H)^{2/5}}
			\right),
\end{equation}
where $F(x,y)$ is a holomorphic function with $F(0,0) = 1$.  For small
$x$ and $y$, it has to behave as $F(x,y) \sim 1 + x + y$ plus terms with
higher powers in $x$ and $y$ to be consistent with a perturbation in
terms of $\lambda$ and $M$.  However, higher powers in $x$ and $y$ lead
to a singular behavior when $\Lambda \rightarrow 0$ (weakly coupled
limit), and are not allowed.  Therefore, $F(x,y) = 1 + x + y$ {\it
exactly},\/ and hence $W_{\rm total} = W_{\rm n.p.} + W_{\rm tree}$.

\begin{figure}
\centerline{\psfig{file=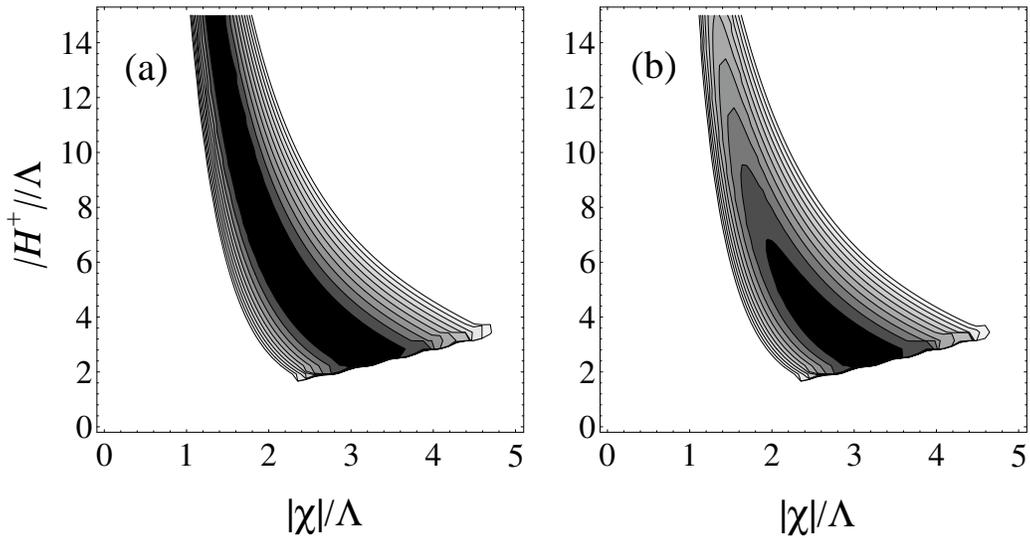,width=\textwidth,angle=-90}}
\caption[1]{Contour plots of the potential in the SO(10) model.  See the
text for the definition of the fields $\chi$ and $H^+$.  We eliminated
$H^-$ using the $D$-flatness condition Eq.~(\ref{D-flat}).  Darker
region has lower energy.  The white region has a potential energy larger than
$0.1 \Lambda^4$.  The sharp cutoff from below is due to the $D$-term
constraint $|H^+|^2 \geq |\chi|^2/2$.  The choice of the parameters is
(a) $(M,\lambda) = (0, 0.01)$, (b) $(M,\lambda)=(0.01 \Lambda,0.01)$.
The vacuum energy is $V \simeq 6\times 10^{-4} \Lambda^4$ in case (b). }
\end{figure}

We first discuss the case where $H$ is massless, {\it i.e.}\/, $M=0$.
In this case, there is a moduli space of the supersymmetric vacua
defined by\footnote{The moduli space can be parametrized by the
gauge-invariant superfield $H^2$.  Note that $U(1)_M^3$ anomaly is
saturated by $H^2$: $2^3 \times 10 + (-1)^3 \times 16 = 64 = 4^3$.
$U(1)_R$ symmetry is explicitly broken by $\lambda \neq 0$ and hence
does not give us useful constraints on the low-energy particle
content.  Even though one should be able to discuss the dynamics with
this composite field $H^2$, we prefer to use the elementary fields
$H^\pm$ and $\chi$ as in \cite{ADS} because the K\"ahler potential has a
much simpler form.  Such a treatment is valid when both $\lambda$ and
$M$ are small.}
\begin{equation}
\psi \psi H = \sqrt{2} \chi^2 H^+ =
\left( \frac{2c}{5\lambda} \right)^{5/7} \Lambda^3 . \label{flat}
\end{equation}
When $\lambda$ is small, the moduli space is far away from the origin,
and one can study the potential explicitly with perturbative K\"ahler
potential.  See Fig.~1(a) for the moduli space Eq.~(\ref{flat}) which
extends to the infinity $H^+ \rightarrow \infty$.
Note that the $D$-flatness requires $|H^+|^2 - |\chi|^2/2
= |H^-|^2 \geq 0$ and the moduli space does not extend to the the region
$|H^+|^2 < |\chi|^2/2$.

Now we turn on the mass term $\frac{1}{2} M H^2 = M H^+ H^-$.  The
conditions for the supersymmetric vacua are
\begin{eqnarray}
\left( - \frac{2}{5}
	\frac{\Lambda^{21/5}}{(\sqrt{2}H^+ \chi^2)^{7/5}}
	+ \lambda \right) \chi H^+ = 0, \label{1} \\
\left( - \frac{2}{5}
        \frac{\Lambda^{21/5}}{(\sqrt{2}H^+ \chi^2)^{7/5}}
        + \lambda \right) \sqrt{2} \chi^2
	+ M H^- = 0, \label{2} \\
M H^+ = 0, \label{3}
\end{eqnarray}
along with the $D$-flatness condition Eq.~(\ref{D-flat}).  It is easy to
see that the supersymmetry is spontaneously broken once there is
non-vanishing $M$ as follows.  Eq.~(\ref{3}) requires $H^+ = 0$, and
then the $D$-flatness requires $H^- = \chi = 0$.  But this is
inconsistent with Eqs.~(\ref{1}), (\ref{2}).  Therefore, any finite $M$
breaks supersymmetry dynamically.  For small $M$, the flat direction in
Eq.~(\ref{flat}) is still almost flat, with $M$ raising the potential
for larger $H^+$. The minimum lies along the direction Eq.~(\ref{flat})
with smallest possible $H^+$, and hence $H^- = 0$, $|H^+| =
|\chi|/\sqrt{2}$ (see Fig.~1(b)).  There is a unique ground state with
no massless scalars.\footnote{Since $U(1)_R$ is explicitly broken by
$\lambda \neq 0$, there is no $R$-axion in this case, similar to the
model with a massive singlet in \cite{NS}.} For larger $M$, the vacuum
is pulled towards smaller values of $H^+$ and $\chi$.  We studied
numerically that the vacuum energy becomes larger for larger $M$.

We can never study the limit $M \rightarrow \infty$ exactly using the
elementary fields because one enters into an intrinsically strongly
interacting regime.  In fact, all the field values approach $\sim
\Lambda$ from above as we gradually raise $M$, and we lose our handle on
the K\"ahler potential.  However, this analysis does show that the
Witten index of the SO(10) model with a {\bf 16} and {\bf 10} vanishes
for any small but finite values of $M$.\footnote{Note also that the
superpotential $W_{\rm total}$ breaks supersymmetry if we regard $\psi
\psi H$ and $H^2$ as true degrees of freedom, since $\partial W/\partial
(H^2) = M/2 \neq 0$.  However one needs to discuss the singularities in
the K\"ahler potential to justify this argument.}
Assuming the continuity of the phase as we gradually increase $M$,
we obtain vanishing Witten index for
$M \gg \Lambda$, which is equivalent to an SO(10) model with a single
{\bf 16}.  If the Witten index of a model vanishes, the model
generically breaks supersymmetry, even though one cannot logically
exclude the possibility of having equal number of supersymmetric
zero-energy states for both bosonic and fermionic states.  Therefore, we
confirm the conclusion in Ref.~\cite{SO10} that the SO(10) model with a
single {\bf 16} breaks supersymmetry dynamically.\footnote{Another
interesting point is that this analyis confirms the spontaneous
breakdown of U(1)$_R$ symmetry in this limit as conjectured in
\cite{SU5,SO10}.  If we take $\lambda = 0$, there is an exact U(1)$_R$
symmetry.  Having $M\neq 0$ leads to a well-defined vacuum with
dynamical supersymmetry breaking.  One can easily see that U(1)$_R$
symmetry is broken spontaneously at this vacuum, and there is an
$R$-axion.}

\begin{table}
\centerline{
\begin{tabular}{c|ccc|ccc|ccc}
& $H$ & $\psi$ & $(\phi,\bar{H})$ &
$\psi\phi\bar{H}$ & $\psi\psi H$ & $(\bar{H}H,\phi H)$ & $f$ &$h$ & $M$
\\ \hline
U(1)$_R$ & 8 & 0 & $-$6 & $-$12 & 8 & 2 & 14 & $-$6 & 0 \\
U(1)$_M$ & $-$4 & 2 & $-$1 & 0 & 0 & $-$5 & 0 & 0 & 5 \\
U(1)$_Y$ & 3 & 1 & $-$3 & $-$5 & 5 & 0 & 5 & $-$5 & 0 \\
SU(2) & 1 & 1 & 2 & 1 & 1 & 2 & 1 & 1 & 2
\end{tabular}
}
\caption{Charges of the fields and parameters under non-anomalous global
symmetries in the SU(5) model with $H$({\bf 5}), $\psi$({\bf 10}),
$(\phi, \bar{H})$({\bf 5}$^*$).}
\end{table}

\setcounter{footnote}{0}
A similar analysis can be done for the SU(5) model with $\psi$({\bf 10})
and $\phi$({\bf 5}$^*$).  Again we introduce vector-like fields $H$({\bf
5}) and $\bar{H}$({\bf 5}$^*$).  There are flat directions which can be
parametrized by four chiral gauge-invariant fields $\psi \phi \bar{H}$,
$\psi \psi H$, $\bar{H} H$, $\phi H$.  The low energy theory along the
flat directions is a pure SU(2) Yang--Mills theory with four singlet
chiral superfields.  The exact superpotential including the
non-perturbative effects is\footnote{We can always make an SU(2)
rotation between $\bar{H}$ and $\phi$ to allow mass term only for
$\bar{H}$.  All other terms in $W$ are invariant under this SU(2) rotation.}
\begin{equation}
W = c \frac{\Lambda^6}{(\psi\psi H)^{1/2} (\psi\phi\bar{H})^{1/2}}
	+ h \psi\psi H + f \psi \phi \bar{H} + M H \bar{H} .
\end{equation}
In the limit of $M \rightarrow 0$, there is a moduli space of
supersymmetric vacua defined by\footnote{The moduli space is
parametrized by SU(2)-doublet chiral superfield $\bar{H}H$ and $\phi H$.
Both U(1)$_R$ and U(1)$_Y$ are broken explicitly by $f$ and $h$, but
SU(2) and U(1)$_M$ are not broken.  The anomalies match with this
particle content.  U(1)$_M^3$: $(-4)^3\times5 + 2^3\times10 +
(-1)^3\times 5\times 2 = -250 = (-5)^3\times 2$, U(1)$\times$SU(2)$^2$:
$(-1)\times 5 = -5$.}
\begin{eqnarray}
\psi \psi H &=& \sqrt{c} \Lambda^3 \frac{(4fh)^{1/4}}{2h}\ , \\
\psi\phi\bar{H} &=& \sqrt{c} \Lambda^3 \frac{(4fh)^{1/4}}{2f}\ .
\end{eqnarray}
Again, the combination of the $D$-flatness and small but non-vanishing
$M$ leads to the spontaneous breakdown of supersymmetry.

In summary, we proposed a simple method to analyze non-calculable
supersymmetric gauge theories.  In the SO(10) model with a single {\bf
16}, there is no flat direction and it cannot be analyzed with the
holomorphy.  When we introduce a {\bf 10}, the model
has flat directions and can be analyzed unambiguously.  By introducing a
small mass of {\bf 10}, one can show the supersymmetry is spontaneously
broken, and hence  Witten index of this model vanishes.  Assuming the
continuity of the phase, Witten index remains vanishing for larger $M$.
In the limit $M \rightarrow \infty$, the model reduces to the original
SO(10) model with a single {\bf 16}, and is expected to break
supersymmetry.

\vskip\baselineskip
\noindent {\sc Acknowledgements}

We thank John March-Russell and Markus Luty for the collaboration at
early stage of this work.  We also thank Hirosi Ooguri and Christopher
Carone for useful comments and careful reading of the manuscript.
This work was supported by the Director, Office of Energy Research,
Office of High Energy and Nuclear Physics, Division of High Energy
Physics of the U.S. Department of Energy under Contract DE-AC03-76SF00098.

\vskip\baselineskip
\noindent {\sc Appendix}
\appendix
This appendix summarizes the notation, and shows the flat direction
explicitly.  It is known that the gamma matrices of SO(8) can be chosen
real, $\gamma^1$, \ldots $\gamma^8$, and $\gamma^9$ is also real in this
basis (this is why there is Majorana--Weyl spinor in SO(8)).  We define
the SO(10) gamma matrices by
\begin{eqnarray*}
\Gamma^1 &=& \gamma^1 \otimes \sigma^1 \\
&\vdots& \\
\Gamma^8 &=& \gamma^8 \otimes \sigma^1 \\
\Gamma^9 &=& \gamma^9 \otimes \sigma^1 \\
\Gamma^{10} &=& 1 \otimes \sigma^2 \\
\Gamma^{11} &=& 1 \otimes \sigma^3 .
\end{eqnarray*}
The Weyl spinor $\psi$({\bf 16}) is defined by $\Gamma^{11} \psi = +
\psi$, and hence can be written as $\psi = \tilde{\psi} \otimes
\uparrow$ with $\tilde{\psi}$ having 16 comonents while $\psi$ has 32
components.  The charge conjugation matrix $C$ in SO(10) spinor is
nothing but $C = \Gamma^{10}$ in this basis.

Without a loss of generality, we can always make an SO(10) rotation to
bring $H$({\bf 10}) to the form
\begin{equation}
H = (0, 0, 0, 0, 0, 0, 0, 0, H^9, H^{10}).
\end{equation}
Then all $D$-terms vanish except that for the rotation between 9th and
10th components.  This leaves SO(8) gauge symmetry unbroken.

Since SO(8) spinors are real, the contribution of $\psi$ to the $D$-terms for
SO(8) generators vanishes when one takes $\tilde{\psi}^* = \tilde{\psi}$
up to a phase.  Note that the generators of SO(8) rotations commute with
$\gamma^9$ while those of
$(i,9)$ and $(i,10)$ rotations anti-commute for $i=1,\ldots,8$.  Therefore, all
$D$-terms vanish for $(i,9)$ and $(i,10)$ rotations if all the
non-vanishing components in $\tilde{\psi}$ have the same chirality
under $\gamma^9$, and we take $\gamma^9 \tilde{\psi} = + \tilde{\psi}$.
For convenience, we fix our basis such that $\gamma^9 = \sigma^3 \otimes
\sigma^3 \otimes \sigma^3 \otimes \sigma^3$.
Recall SO(8) real spinor is equivalent to a vector representation up to
an outer automorphism
(triality), and we can take $\tilde{\psi} = (\uparrow \otimes \uparrow
\otimes \uparrow \otimes \uparrow) \chi$ using an SO(8) rotation without a
loss of generality.  The unbroken symmetry is SO(7).

Now the only $D$-term which we have to discuss is that of the (9,10)
rotation.  It is convenient to define $H^\pm = (H^{10} \mp i
H^9)/\sqrt{2}$ so that $H^\pm$ have eigenvalues $\pm 1$ under (9,10)
rotation.  $\chi$ has an eigenvalue $-1/2$.  Therefore the $D$-flatness
requires
\begin{equation}
|H^+|^2 - |H^-|^2 - \frac{1}{2} |\chi|^2 = 0,
\end{equation}
as in Eq.~(\ref{D-flat}).  The gauge invariant fields are
\begin{eqnarray}
\psi \psi H &\equiv& {}^t \psi C \Gamma^\mu \psi H^\mu \nonumber \\
	&=& {}^t \tilde{\psi} \tilde{\psi} H^{10}
		+ {}^t \tilde{\psi} (- i \gamma^i) \tilde{\psi} H^i
		\nonumber \\
	&=& \sqrt{2} \chi^2 H^+, \\
H^2 &\equiv& H^\mu H^\mu \nonumber \\ &=& 2 H^+ H^-
\end{eqnarray}
for $\mu = 1, \ldots, 10$ and $i = 1, \ldots, 9$.

\end{document}